\documentclass[12pt]{article} 
\usepackage{epsfig,amssymb,wasysym} 
\def\beq{\begin{equation}}
\def\eeq#1{\label{#1}\end{equation}}
\def\eeqn{\end{equation}}

\def\beqa{\begin{eqnarray}}
\def\eeqa#1{\label{#1}\end{eqnarray}}
\def\eeqan{\end{eqnarray}}

\let\bar=\overbar

\def\Dslash{\not{\hbox{\kern-4pt $D$}}}
\def\dslash{\not{\hbox{\kern-2pt $\del$}}}

\def\msb{{\bar{\ssstyle M \kern -1pt S}}}

\def\Title#1{\begin{center} {\Large {\bf #1} } \end{center}}
\begin{document} 
\Title{OPERA: waiting for the $\tau$}
\bigskip\bigskip 
\begin{raggedright}  
{\it Andrea Longhin\index{Longhin, A.}\footnote{on behalf of the OPERA Collaboration}\\ 
Physics Department ``M. Merlin'',\\ 
Bari University\\ 
via Amendola 173,\\ 
I-70126 Bari, Italy}\\
\bigskip \bigskip
\end{raggedright}
%
%
\bigskip \bigskip 
\section{Overview and physics reach}

OPERA\cite{1.} is a long baseline (730 Km) neutrino
oscillation experiment located in Italy at the Gran Sasso underground laboratory
(3500 m.w.e. overburden, residual $\mu$ flux $\sim$ 1 h$^{-1}$m$^{-2}$).
The detector is conceived to observe $\nu_\mu \to \nu_\tau$ oscillations in the
parameter region indicated by Super-Kamiokande\cite{2.}\footnote{
and confirmed by K2K and MINOS, not to speak about Kamiokande, SOUDAN-2 and MACRO.} through direct
observation (appearance) of $\nu_\tau$  
in an almost pure $\nu_\mu$ beam (contaminations: $\sim$ 2 \% $\bar{\nu}_\mu$, $\lesssim$ 1\%
$\nu_e+\bar{\nu}_e$ and negligible $\nu_\tau$).
The CERN Neutrinos to Gran Sasso (CNGS \cite{3.}) high energy neutrino beam
($\langle E_{\nu_{\mu}}\rangle \simeq 17$ GeV) has been designed in order to
maximize the possible number of $\nu_\tau$ charged current interactions 
at destination taking into account the energy dependance of the oscillation
probability and the $\tau$ production cross section.

The OPERA detector is a massive (1.35 kton) and highly modular
lead-nuclear emulsion target composed of 154750 units called Emulsion Cloud
Chambers (ECCs or ``bricks''). Each brick is a 56-layer stack of lead plates interleaved with
nuclear emulsions providing the $\mu$m and the mrad level precision tracking needed for
detecting the $\tau$ decay topology.

At CNGS energies the average $\tau$ decay length is $\sim$ 450 $\mu m$.
$\nu_\tau$ appearance will be identified by the detection of the peculiar $\tau$
lepton decay topology through its decay modes into electron (17.8\%), muon
(17.7\%), and single (50\%) or three charged hadrons (14\%).
In the case of a decay in the same lead plate of production, the impact parameter
of the daughter track with respect to primary vertex  can be used while for
longer decays in which the $\tau$ traverses at least one emulsion layer, the
kink angle in space between the charged decay daughter and the parent direction
will be employed.

Each one of the two targets is instrumented by 31 planes of electronic detectors
(horizontal and vertical arrays of 2.6$\times$1 cm thick scintillator strips
read by WLS fibres and multi-anode PMT at both ends) that allow the location
of the brick in which the interaction occurred and drive the scanning of the
emulsions by providing information on the outgoing tracks. The trigger efficiency 
is as large as 99\%. 

A magnetic spectrometer follows the instrumented target and 
measures the charge and momentum of penetrating tracks. Each spectrometer is composed
by a bipolar iron magnet ($\sim$ 990 tons, B = 1.52 T) instrumented with 22 RPC planes 
($\sim$ 70 m$^2$ each) which act as inner trackers and six fourfold drift tubes 
(8 m long) planes which provide high precision tracking with a point resolution better 
than 300 $\mu$m. Precise charge measurement is particularly important for the efficient 
suppression of the charm background. A resolution $\Delta p / p < 0.25$ and charge 
mis-identification of a few $\permil$ up to $\sim$ 25 GeV can be obtained. 

The $\tau$ search sensitivity calculated for 5 years of data taking with a
total number of 4.5 $\cdot$ 10$^{19}$ integrated p.o.t./year (200 days runs) 
is given in Table \ref{tab:ntau}. The number of signal events essentially scales 
like $({\Delta m^2_{13}})^2$.

The main background sources are given by large angle scattering of muons
produced in ordinary charged current interactions, hadronic interactions of daughter
particles produced at primary interaction vertex and prompt charmed particles
decays associated with inefficiency on the primary muon identification.

Figure \ref{tab:ntau} shows the probability of discovery at 3 and 4 $\sigma$
significance as a function of $\Delta m^2_{13}$. 

\begin{table}[hbt!]
\begin{minipage}[c]{8cm} 
\begin{center} 
{\small{ \begin{tabular}{|l|c|c|c|}
\hline
&Signal&Signal&Bckg\\ &{\tiny{$\Delta m^2_{13}=2.5\cdot 10^{-3} eV^2$}}&{\tiny{$\Delta m^2_{13}= 3.0 \cdot 10^{-3} eV^2$}}&
\\ \hline $\tau \to \mu$ & 2.9 & 4.2 & 0.17\\ \hline $\tau \to e$ & 3.5 & 5.0 &
0.17\\ \hline $\tau \to h$ & 3.1 & 4.4 & 0.24\\ \hline $\tau \to 3h$ & 0.9 & 1.3
& 0.17\\ \hline $ALL$ & 10.4 & 15.0 & 0.76\\ \hline \end{tabular} }}
\end{center} 
\end{minipage}%
\begin{minipage}[c]{1cm}
\hspace{0.5cm}
\end{minipage}%
\begin{minipage}[r]{8cm} \begin{center}
\epsfig{file=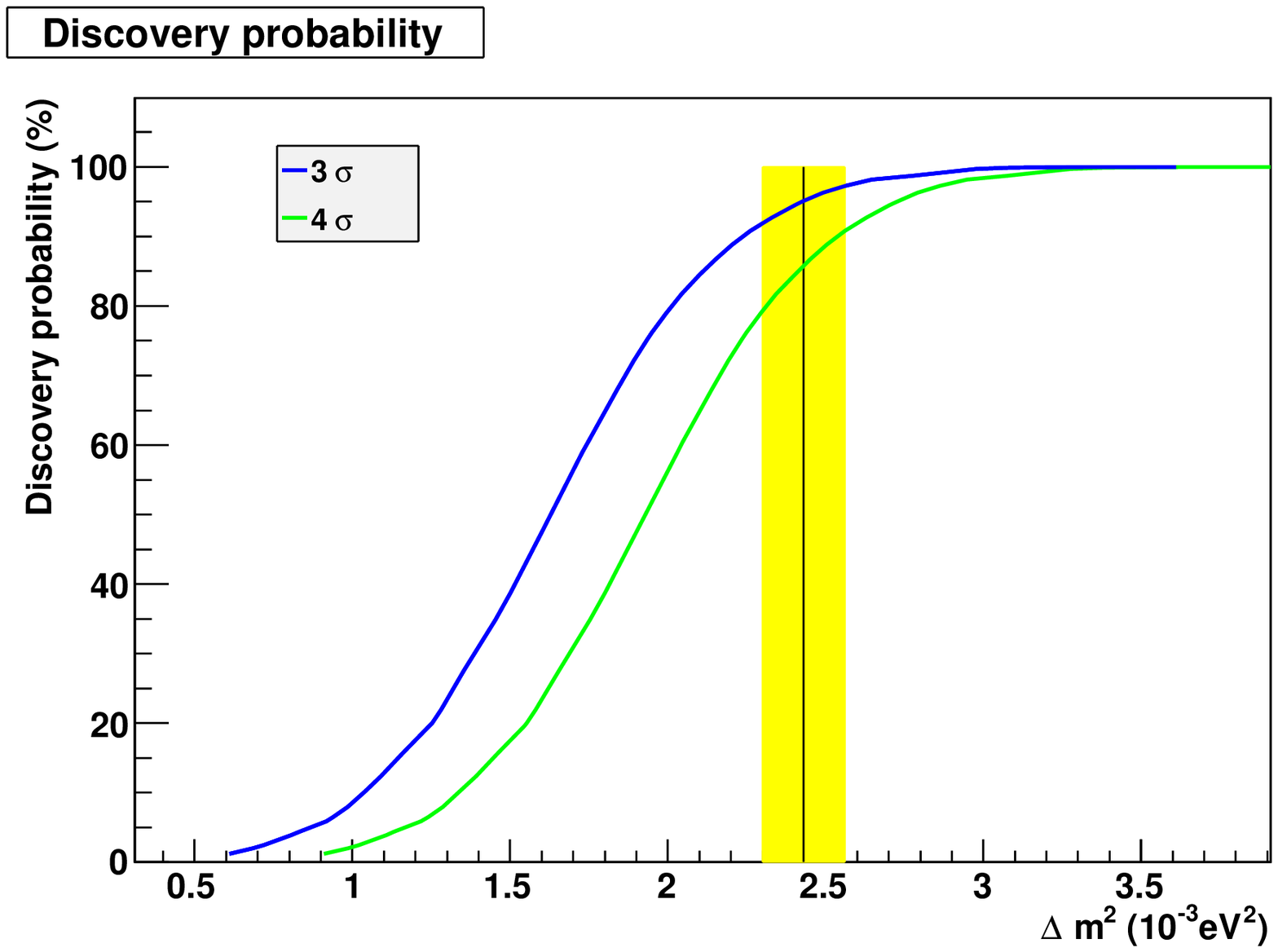,height=1.8in} 
\label{fig:discovery} 
\end{center}
\end{minipage} \caption{See text. The shaded band in the plot marks the region
indicated by global analysis after the recent MINOS determination.}
\label{tab:ntau} \end{table}

The OPERA proposal dates back to 2000, construction started in mid-2003 and the
electronics part was completed by the first half of 2007. Detector filling with 
bricks started in 2007 and was completed by mid-2008.

\section{Data taking and results}

A summary of the first data taking of OPERA is shown in Table \ref{tab:runs}.

\begin{table}[hbpt!] 
\begin{center} 
\begin{tabular}{|c|c|c|c|c|}  
\hline 
Period & ($10^{18}$ p.o.t.) & Filling &Events (target/external) & Comment \\ \hline 
17-30 Aug 2006 & 0.76 & 0 &0/319&{\scriptsize{elect. det. commissioning}} \\ \hline 
Oct 2006 & 0.06 & 0.1\% & 0/29 & {\scriptsize{final commissioning}} \\ \hline 
Oct 2007 & 0.79 & 40\% &38/331 & {\scriptsize{first events in emulsions}} \\ \hline 
Jul-Oct 2008 & $\sim$10 &100\%& $\sim$1000/3500 & {\scriptsize{regular running}}\\ \hline 
\end{tabular} 
\caption{Data taking phases and collected statistics. Full details are given in the text.}
\label{tab:runs} 
\end{center} 
\end{table}
The first CNGS technical run started in August 2006. Since brick filling had not
started yet, this run was dedicated to the commissioning of the electronic
detectors and to alignment and reconstruction algorithms tuning.
A sample of 319 neutrino-induced events was collected coming from interactions
in the rock surrounding the detector, in the spectrometers and in the target
walls. 
Fine-tuning of the synchronization between CERN and Gran Sasso, performed using GPS clocks, was also possible. The beam spill timing structure composed by two 10.5 $\mu$s wide
bunches separated by 50 ms could be clearly observed. The zenith-angle
distribution of the muon tracks associated to beam $\nu_\mu$ interactions in the rock 
was measured to be centered at 3.4 $\pm$
0.3(stat.)$^\circ$, in agreement with the value of 3.3$^\circ$ expected from
geodesy. Finally, using a Monte Carlo simulation tuned on data from the MACRO
experiment, angular shape and absolute normalization of cosmic muons could be
reproduced \cite{OPERArun2006}.

In October 2006, a new run began but was shortly interrupted (0.06$\cdot$10$^{18}$ p.o.t.) due to a leak in the closed water cooling system of the reflector in the neutrino beam line.

After repair, a new physics run was possible in October 2007, when OPERA had 40\% of the
target mass installed (about 550 tons). The beam extraction intensity was
limited to 70\% of the nominal value due to beam losses which brought severe
radiation damage to the CNGS ventilation control electronics. In about 4 days of
continuous data taking, 0.79 $\cdot$ $10^{18}$ p.o.t. were delivered and 38
neutrino interactions in the OPERA target were triggered by the electronic
detectors with an expectation of $31.5 \pm 6$. Out of these 29 had
charged-current and 9 neutral-current topology. Out-of-target interactions
amounted to 331 events to be compared with an expectation of 303. 

The 2007 run provided the opportunity for the first test on real neutrino
interactions for the complex chain of brick location, validation, handling,
emulsion gridding, development and finally automatic scanning. 

The essential interplay between electronic detectors and emulsions could 
also be carefully tested profiting of this initial sample.
Position of bricks obtained from extensive alignment measurements and mechanical
model of structure deformation allowed an effective brick finding of 80 $\pm$ 7\%.
Wall finding efficiency was greater than 95\% despite the frequent emission of 
low momentum back-scattered charged particles.

A key tool for brick finding is the Changeable Sheet doublet (CS)\cite{9.}, 
consisting of a pair of emulsion films hosted in a box placed outside the brick which 
acts as an interface between the brick and the electronic target tracker. 

The positive finding of tracks compatible with the electronic detector 
predictions in the CS doublet validates the brick finding algorithm prediction.
Following the need for high purity, before installation the CS emulsions underwent 
a specific treatment called ``refreshing'' 
(a period during which storage in a high humidity and temperature environment is applied 
which allows to ``erase'' previously recorded tracks). 
The CS refreshing and assembling of the doublets was done underground to avoid
contamination from cosmic tracks
\footnote{Refreshing was indeed performed for all emulsions underground (Tono mine) in Japan before 
their shipment to Italy but, in the case of the large sample of brick emulsions, it was not
repeated in Gran Sasso. This was a viable strategy thanks to the fact that the presence of tracks recorded during transportation can be easily dealt with in this case.}.
In case of positive validation by the CS the brick is brought to surface and exposed 
to cosmic rays\footnote{This is done in a dedicated area with a properly designed shielding 
which is intended to suppress the low energy component.} before development for plate-to-plate 
fine alignment. Before detaching the CS from the brick, they are exposed to four thin X-ray beams, 
in order to define their relative alignment. 

Among the 38 bricks, 36 had a good CS tagging. The measured residuals between electronic 
detectors predictions and CS tracks were found to be of the order of a few cm.
CS to brick connection was achieved with 54 $\mu$m and 9 mrad position and slope accuracy.

The emulsions of the selected bricks were sent to the various automated scanning microscopes spread throughout Europe and Japan (about 40).
All the tracks located in the CS were subsequently followed upstream inside the brick 
(scan-back) up to a ``stopping point''. A general scanning (no angular preselection) was subsequently performed in a volume around the stopping point(s) in order to reconstruct the vertex topology.
The mechanical accuracy obtained during the brick piling is in the range of 50-100 $\mu$m. The reconstruction of cosmic rays passing through the whole brick allows to improve the definition of a global reference frame, leading the precision to about 1-2 $\mu$m. 
The technique of marking emulsions with thin lateral X beams to get fast
alignment pattern to be used in tracks scan-back and CS internal alignment with
Compton tracks have been also successfully applied \cite{9.}.

\begin{figure} 
\begin{center}
\begin{tabular}{ll}
a)~~
{\epsfig{file=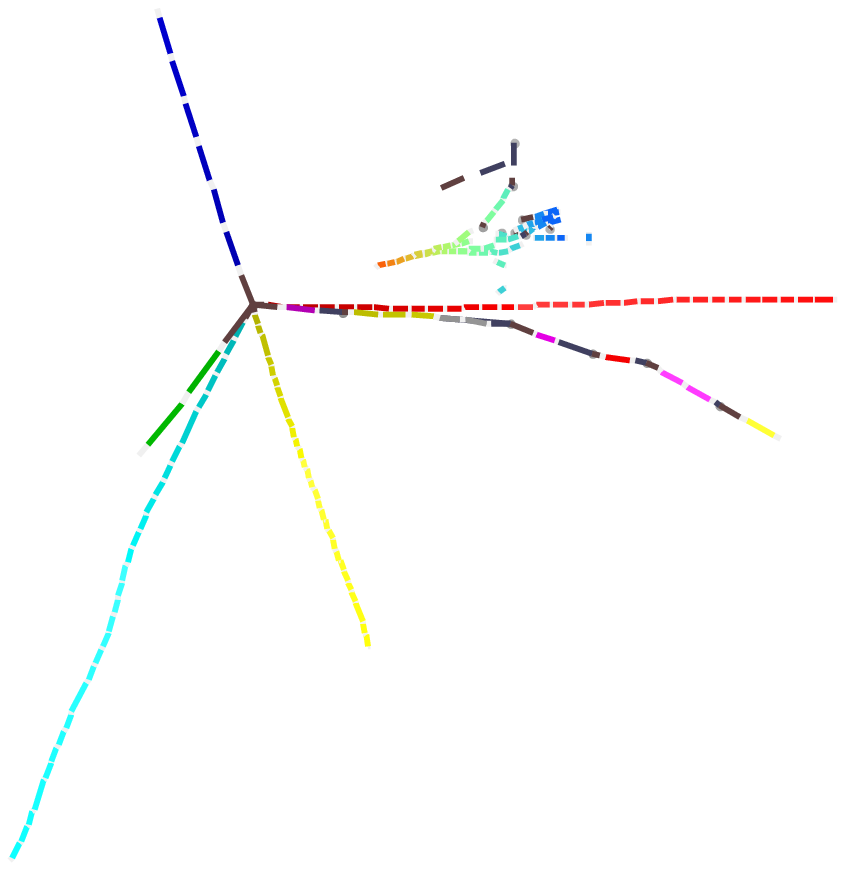,height=1.5in}}
{\epsfig{file=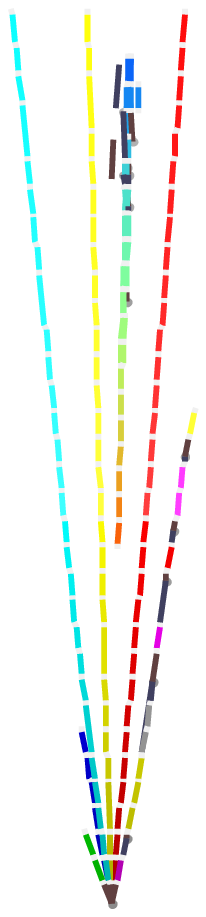,height=1.8in}}
{\epsfig{file=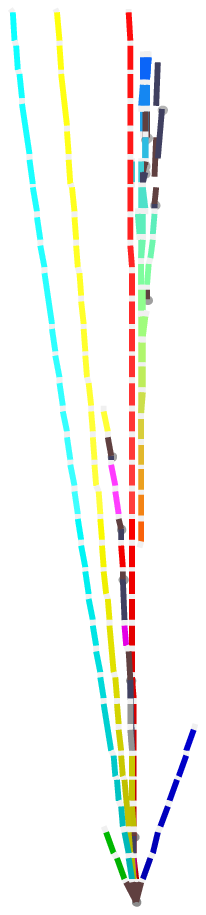,height=1.8in}}&
b)~~\epsfig{file=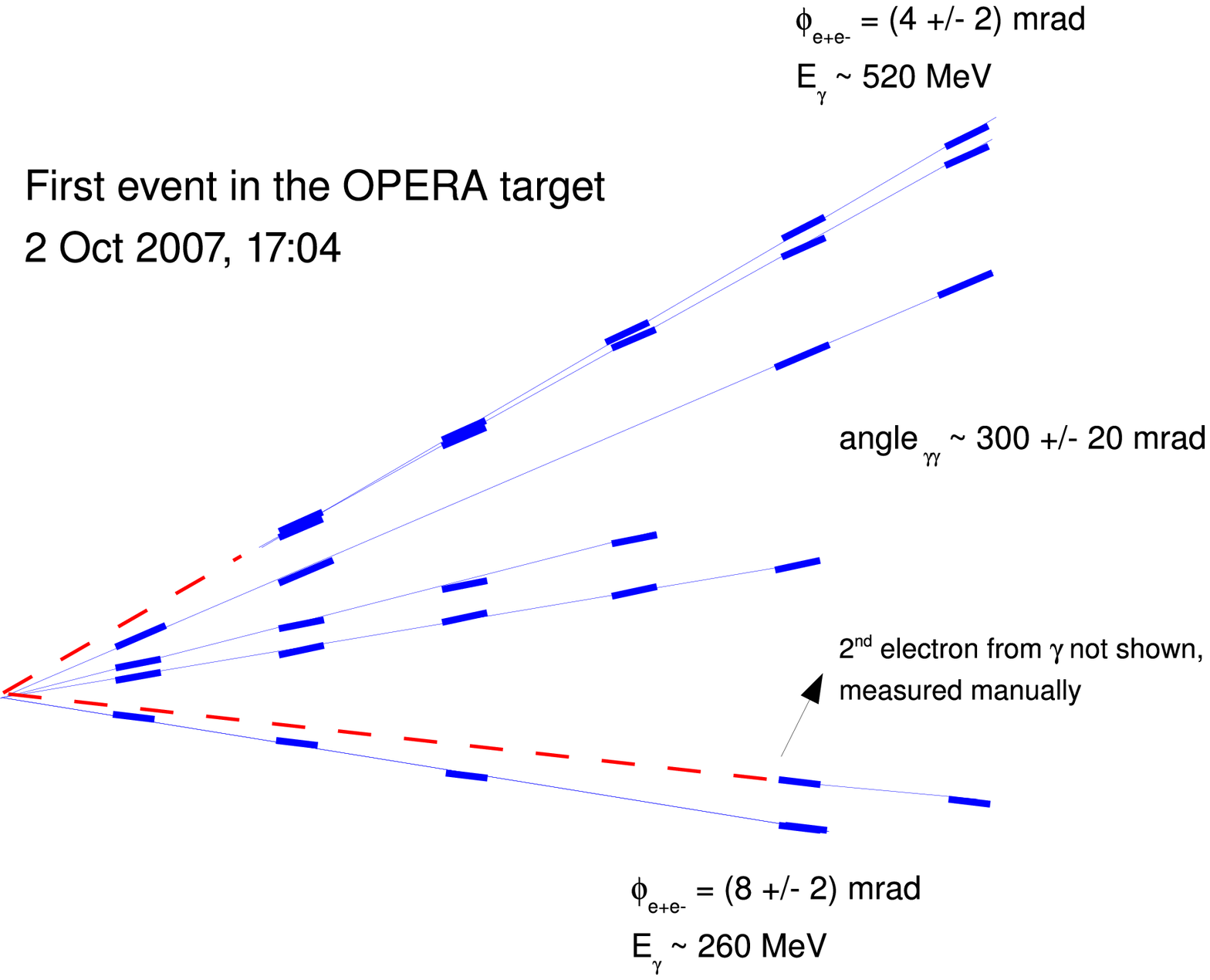,height=2.2in} \\
\end{tabular}
\caption{
Displays of two $\nu_\mu^{CC}$ neutrino vertices from the 2007 run reconstructed in the brick. 
Segments represent emulsion tracks ($\sim$ 300 $\mu$m thick), gaps are due to 1 mm thick lead plates.
a) the frontal and two orthogonal lateral views are shown. b) thick dashed lines represent the trajectory
of $\gamma$s before conversion.}
\label{fig:events} 
\end{center} 
\end{figure}

In Figure \ref{fig:events} the display of two $\nu_\mu^{CC}$ vertices reconstructed in the brick is shown. The first one is an interaction with 6-prongs and an electromagnetic shower pointing to the primary interaction vertex.
In the second 4-prong vertex a decay of a $\pi^0\to \gamma(\to e^+e^-)\gamma(\to e^+e^-)$ has been fully reconstructed (with a thickness of 7.8 cm a brick amounts to about 10 $X_0$). The kinematic analysis 
leads to an invariant mass measurement $m_{\gamma\gamma} = (110\pm 30)$ MeV.

\begin{figure} 
\begin{center}
\begin{tabular}{ll}
a)~~\epsfig{file=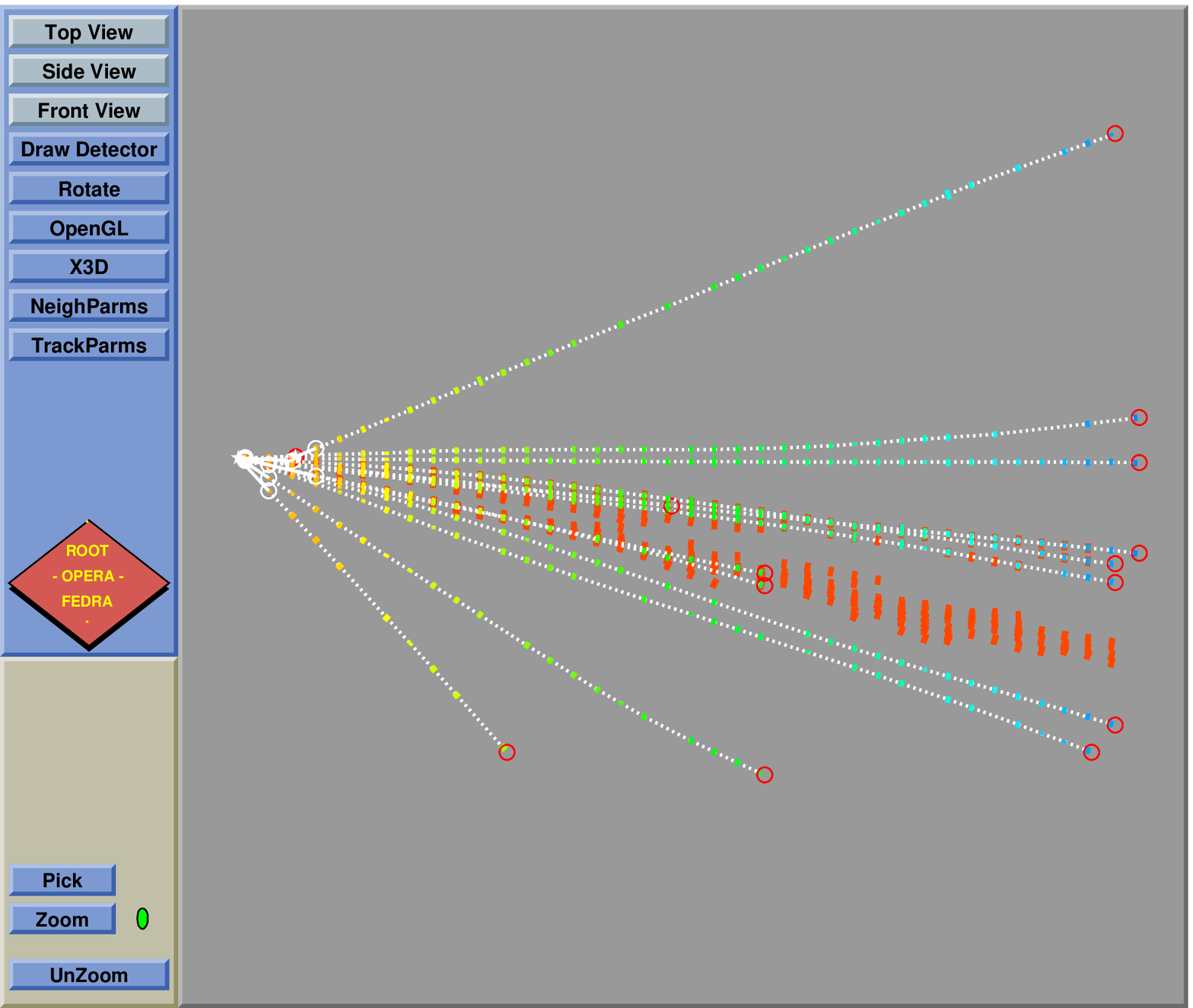,bbllx=88pt,bblly=0pt,bburx=567pt,bbury=479pt,height=2.2in,width=2.6in,clip=}&
b)~~\epsfig{file=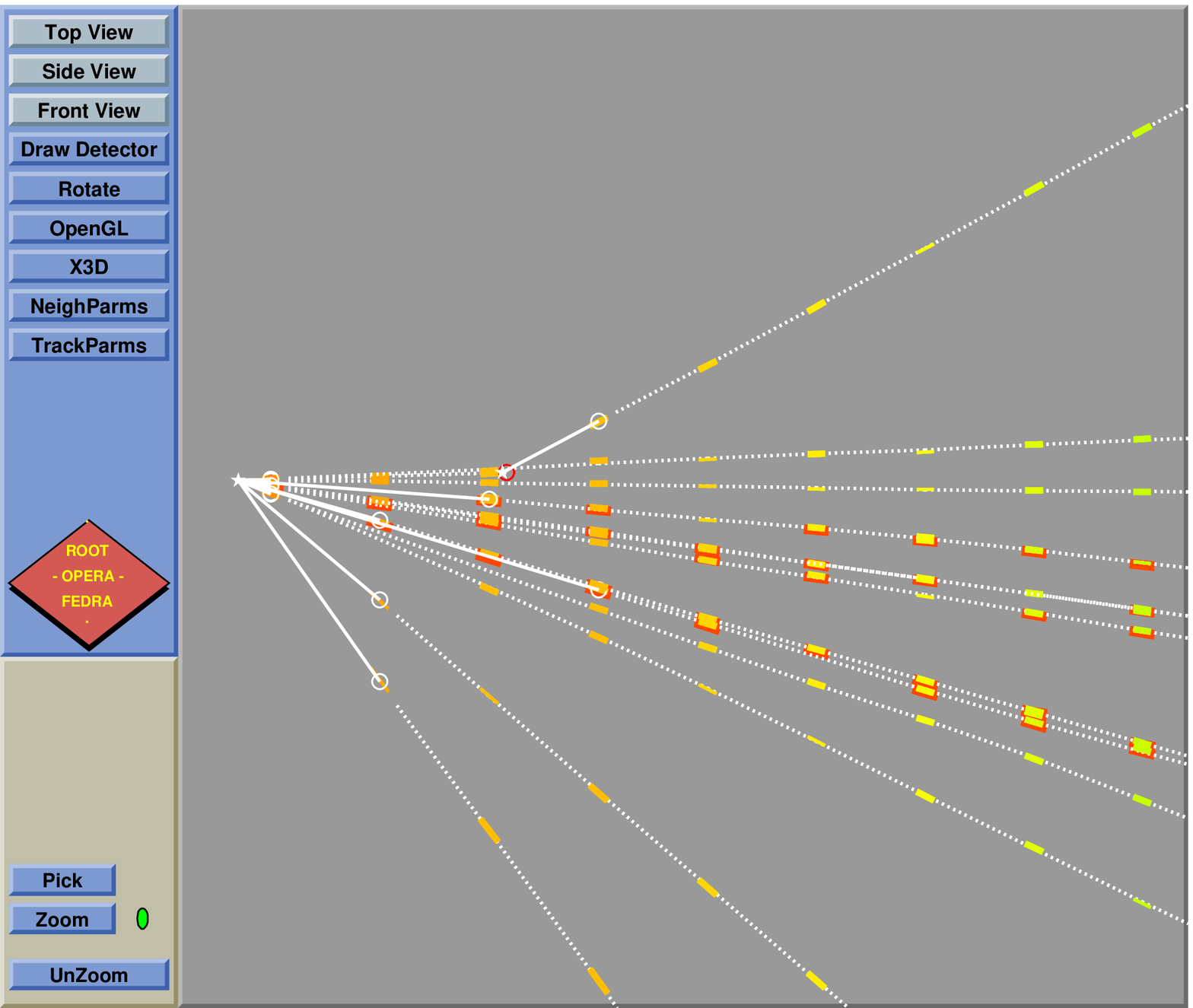,bbllx=88pt,bblly=0pt,bburx=567pt,bbury=479pt,height=2.2in,width=2.6in,clip=}\\
\end{tabular}
\caption{
Displays of a charm candidate neutrino vertex in the brick from the 2007 run. 
Segments represent emulsion tracks ($\sim$ 300 $\mu$m thick), gaps are due to 1 mm thick lead plates.
a) the full brick infomation. b) a zoomed view in the primary vertex region.}
\label{fig:eventsC} 
\end{center} 
\end{figure}

Figure \ref{fig:eventsC} shows the first observed charm candidate. A single prong 
decay topology is visible. The measured kink decay angle is 0.204 rad and the decay length 
is 3247 $\mu$m. The estimated momentum of the daughter track is 3.9$^{+1.7}_{-0.9}$ GeV 
($p_T$ = 0.796 GeV). 
The muon measured by electronic detectors is unambigously matched to the primary vertex
and lies in a back-to-back configuration in the trasverse plane (not shown) with respect to the charmed
hadron candidate and fragmentation tracks as expected. An electromagnetic shower is also visible in the display. 
The observation of one candidate in the sample is statistically in agreement with expectations.

For some selected events tracks from primary vertices were also measured in the adjacent
downstream brick thus validating the connection procedure which is of great importance when
a detailed kinematic reconstruction of the event is required (mainly through momentum measurement 
by multiple Coulomb scattering).

\section{Conclusion and future perspectives}

A major revision of the CNGS project has been taken in the beginning of 2008 in order
to improve the radiation shielding of the electronics and reduce the beam
losses. 
Meanwhile the OPERA target has been completed by early July 2008 in correspondence with 
the start of a new long run of CNGS beam. On 1st of October 2008, 1.0 $\cdot$ 10$^{19}$
p.o.t. have been integrated. Analysis is in progress at the time of writing.
The collected sample amounts to about 1000 neutrino interactions 
of which 750 are expected to be $\nu_\mu^{CC}$ events, 225 $\nu_\mu^{NC}$ events, 42 charm decays, 
6 $\nu_e$ or $\bar{\nu}_e$ events and 0.5 $\nu_\tau$ events. 

The concept of the OPERA experiment has been experimentally validated by
measuring neutrino events in the detector. Using the charm sample the
capability to efficiently reconstruct $\tau$ decays will be fully exploited. 
With some dose of luck the first $\tau$ candidate event could be observed 
in the data from the current 2008 run.

\end{document}